\documentclass[aps,prd,preprint]{revtex4-1}
\usepackage[mathlines]{lineno}
\usepackage{units}
\usepackage{amsmath,amssymb,amsthm,amstext,amsfonts}
\usepackage{graphicx}
\usepackage[pdftex,plainpages=false,pdfpagelabels]{hyperref}
\usepackage{subfigure}

\begin{document}

\title{Photon decay in UHE air showers: stringent bound on Lorentz violation}

\author{Fabian Duenkel}
\affiliation{Department Physik, Universit\"at Siegen, 57068 Siegen, Germany}
\author{Marcus Niechciol}
\affiliation{Department Physik, Universit\"at Siegen, 57068 Siegen, Germany}
\author{Markus Risse}
\affiliation{Department Physik, Universit\"at Siegen, 57068 Siegen, Germany}

\begin{abstract} 

In extensive air showers induced by ultra-high-energy (UHE) cosmic rays,
secondary photons are expected to be produced at energies far above those accessible by other means.
It has been shown that the decay of such photons, as possible in certain theories allowing Lorentz violation,
can lead to significant changes of the shower development.
Based on observations of the average depth of the shower maximum $\left<X_\text{max}\right>$,
a stringent bound on Lorentz violation has been placed in a previous work.
Here we include the shower-to-shower fluctuations $\sigma(X_\text{max})$ as an additional observable.
The combined comparison of $\left<X_\text{max}\right>$ and $\sigma(X_\text{max})$
to shower observations allows a much stricter test of the possible decay of UHE photons,
improving the previous bound by a factor of $50$.
\end{abstract}

\maketitle

\section{Introduction}

In current efforts towards a more fundamental theory in particle
physics, deviations from exact Lorentz symmetry may occur (see e.g.~\cite{liberati09a}).
To test possible effects of Lorentz violation (LV), the extremely high energies of cosmic rays and gamma rays have been used
and some of the best limits on LV were obtained (e.g.~\cite{kostelecky11a,klinkhamer08a,klinkhamer08b,klinkhamer08c,klinkhamer17}).

In this article, we focus on isotropic, nonbirefringent LV in the photon sector.
We specialize to the case of a photon velocity larger than the maximum attainable velocity of standard Dirac fermions~\cite{kostelecky02a}
which allows photon decay as a new process.
Specifically, the impact of this type of LV on extensive air showers initiated by cosmic rays in the Earth's atmosphere is exploited, 
with a focus on ultra-high energies (UHE) above $\unit[1]{EeV} = \unit[10^{18}]{eV}$.
This approach was first studied in~\cite{diaz16a}, where an analytical ansatz was used, modifying the well-known Heitler model for electromagnetic
cascades to include LV through photon decay. A big impact on the longitudinal shower development of electromagnetic cascades was found.
Building upon this idea, a full Monte Carlo (MC) ansatz was used in~\cite{klinkhamer17} to study the impact of LV on air showers initiated by primary hadrons. 
For this case of primary hadrons, additionally the modified decay of neutral pions due to LV~\cite{klinkhamer16a} has been taken into account.
Comparing the predictions of the average atmospheric depth of the shower maximum $\left<X_\text{max}\right>$
for air showers with LV to shower observations, a significant limit on LV could be determined.
As had been remarked already in~\cite{klinkhamer17}, a considerable improvement in sensitivity
may be expected by adding further observables.
Here, we extend that previous work by taking into account the shower-to-shower fluctuations $\sigma(X_\text{max})$
as an additional observable. 
As will be shown, much stricter constraints are indeed possible.

The theory background of LV in the context of this study and some relevant aspects of the previous analyses are briefly summarized in Sec.~\ref{sec:theory}. 
The current analysis is presented in Sec.~\ref{sec:analysis},
in particular
the methodology to compare simulations and data in more than one
obervable and the result after application. Sec.~\ref{sec:discussion} contains a discussion and a brief summary.

\section{Theory background and previous bounds}
\label{sec:theory}

A relatively simple extension of standard quantum electrodynamics
(QED) is used, where a single term which breaks Lorentz invariance but preserves CPT and gauge invariance \cite{chadha83a,Colladay:1998fq,kostelecky02a} is added to the Lagrange density:
\begin{linenomath}
\begin{equation}
\begin{split}
\mathcal{L} = &\underbrace{-\frac{1}{4}F^{\mu\nu}F_{\mu\nu} +
  \overline{\psi}\left[\gamma^\mu(i\partial_\mu-eA_\mu)-m\right]\psi}_{\text{standard
  QED}}\\
&\underbrace{-\frac{1}{4}(k_F)_{\mu\nu\rho\sigma}F^{\mu\nu}F^{\rho\sigma}}_{\text{CPT-even
LV term}}.
\end{split}
\label{eq:lv_lagrangian}
\end{equation}
\end{linenomath} 
Natural units ($\hbar =  c = 1$) and the Minkowski metric $g_{\mu\nu}(x) = \eta_{\mu\nu} =   [\text{diag}(+1,-1,-1,-1)]_{\mu\nu}$ are used here.
The added tensor $(k_F)_{\mu\nu\rho\sigma}$ consists of 20 independent components. Ten of these produce birefringence, eight lead to direction-dependent modifications of photon propagation,
and one corresponds to an unobservable double trace that changes the normalization of the photon field. 

The last component causes an isotropic modification of the photon propagation.
Thus, isotropic, nonbirefringent LV in the photon sector is controlled by a single dimensionless parameter $\kappa$ which is related to the fixed tensor $k_F$ in Eq.~(\ref{eq:lv_lagrangian}) in the following way:
\begin{linenomath}
\begin{equation}
{(k_F) ^{\lambda}}_{\mu\lambda\nu} = \frac{\kappa}{2}\left[\text{diag}(3,1,1,1)\right]_{\mu\nu}.
\label{eq:kF}
\end{equation}
\end{linenomath}
Note that the parameter $\kappa$ is often denoted by
$\tilde{\kappa}_\text{tr}$ in the literature, see
e.g.~\cite{kostelecky11a,kostelecky02a,klinkhamer08c}. The phase velocity of the photon is given by
\begin{linenomath}
\begin{equation}
v_\text{ph} = \frac{\omega}{|\vec{k}|} =
\sqrt{\frac{1-\kappa}{1+\kappa}}\ c.
\label{eq:phasevelocity}
\end{equation}
\end{linenomath}
In physical terms, the velocity $c$ corresponds to the maximum attainable velocity of the massive Dirac fermion in Eq.~(\ref{eq:lv_lagrangian}), whereas the phase velocity $v_\text{ph}$ of the photon is smaller (larger) than $c$ for positive (negative) values of $\kappa$. 
Theory (\ref{eq:lv_lagrangian}) is consistent (i.e., causal and unitary)
for $\kappa \in (-1,1]$~\cite{Klinkhamer:2010zs},
and microscopic models exist for both positive ~\cite{Klinkhamer:2010zs,Bernadotte:2006ya} 
and negative~\cite{Klinkhamer:2011ez} values of $\kappa$.

For non-zero values of $\kappa$, certain processes which are forbidden in the conventional, Lorentz-invariant theory become allowed. In this article, we focus on the case $\kappa < 0$, where photons become unstable above the energy threshold
\begin{linenomath}
\begin{equation}
E^\text{th}_\gamma(\kappa) = 2\,m_e\,\sqrt{\frac{1-\kappa}{-2\kappa}} \simeq \frac{2\,m_e}{\sqrt{-2\kappa}},
\label{eq:photonthreshold}
\end{equation}
\end{linenomath}
in which $m_e \simeq \unit[511]{keV}$ is the rest mass of the electron. Photons with an energy above this threshold decay very efficiently into electron-positron pairs.

The photon decay length drops to scales of centimeters and below right above the threshold,
resembling a quasi-instantaneous decay of photons into eletron-positron pairs~\cite{klinkhamer08c,diaz15a}.
Above-threshold photons from astrophysical sources are not able to reach the Earth.
Therefore, terrestrial observations of gamma rays with energies of the order $\unit[100]{TeV}$
from distant sources
were used to impose an initial limit of~\cite{klinkhamer08c,diaz15a}
\begin{linenomath}
\begin{equation}
\kappa > -9 \times 10^{-16} ~~~  \text{($\unit[98]{\%}$ CL)}~.
\label{eq:limit2008}
\end{equation}
\end{linenomath}
Observations of higher-energy photons would improve this limit.
Extensive searches for astrophysical (primary) photons with PeV or EeV energies were conducted,
but so far no unambiguous photon detection could be reported at these energies (see e.g.~\cite{niechciol17a}).

However, in air showers initiated by UHE hadrons in the Earth's atmosphere, photons with energies ${\gg}\unit[100]{TeV}$ are expected to be produced as secondary particles:
in the first interaction of the primary hadron with an atmospheric nucleus, mostly charged and neutral pions are produced.
The charged pions further interact with particles from the atmosphere, producing more pions, while the neutral pions,
in standard physics, rapidly decay into pairs of photons, which in turn trigger electromagnetic sub-showers.
Especially in the start-up phase of the air shower, where the energy of the secondary particles is very high, a modification of
the particles due to LV (e.g.\ the immediate decay of above-threshold photons) can drastically modify the overall development of the air shower~\cite{klinkhamer17,diaz16a}.

For a consistent treatment within the LV-theory considered, also the modification of the decay of the neutral pion into two photons has to be taken into account.
Roughly speaking, neutral pions become stable for energies exceeding $E^\text{th}_{\pi^0} \simeq 132\, E^\text{th}_\gamma$
(for more details, see~\cite{klinkhamer16a}).
Although it was found that the impact on the longitudinal shower development (which we focus on in this work) is minor~\cite{klinkhamer17},
this effect is also included in the present simulations.

\begin{figure}[p]
  \centering
  \subfigure[]{\label{fig:xmax_hadrons}\includegraphics[width=0.75\textwidth,]{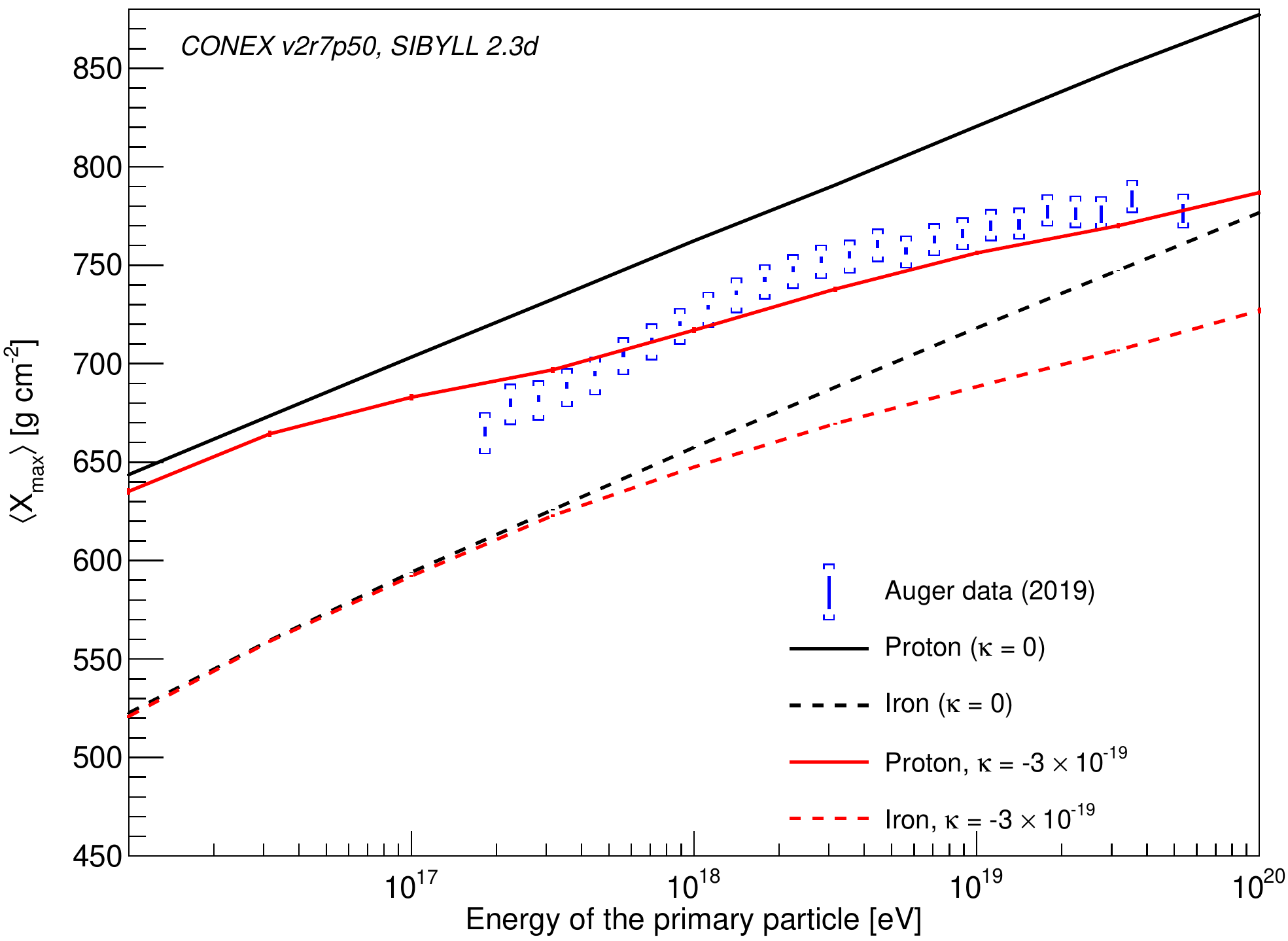}}\\
  \subfigure[]{\label{fig:sigma_hadrons} \includegraphics[width=0.75\textwidth,]{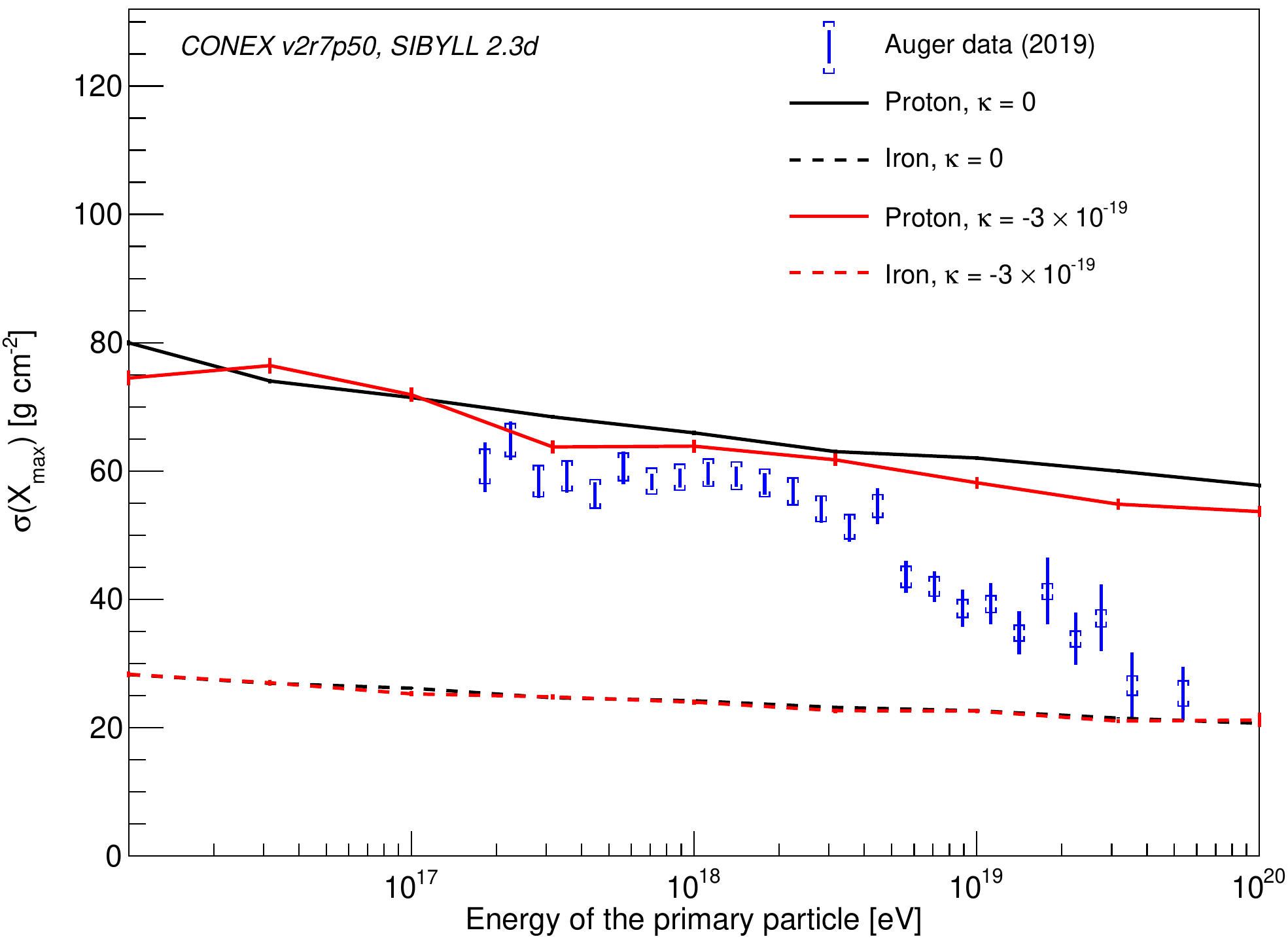}}
\caption{$\left<X_\text{max}\right>$ and $\sigma(X_\text{max})$ as a function of the primary energy for primary protons and iron nuclei for the absence of LV ($\kappa = 0$)
  and for the previous best bound on $\kappa$ (cf.\ Eq.~\ref{eq:limit2017}). Shown are also measurements from the Pierre Auger Observatory~\cite{auger19,auger14a}, with both statistical uncertainties (shown as error bars) and systematic uncertainties (shown as brackets) included.}
\label{fig:xmax_sigma_hadrons} 
\end{figure}

Implementing these modifications in MC simulations of air showers, a strong dependence of the depth of the shower maximum $\left<X_\text{max}\right>$
on $\kappa$ was found \cite{klinkhamer17}, as also displayed in Fig.~\ref{fig:xmax_hadrons}.
Comparing to data, a limit of
\begin{linenomath}
\begin{equation}
\kappa > -3 \times 10^{-19} ~~~  \text{($\unit[98]{\%}$ CL)}
\label{eq:limit2017}
\end{equation}
\end{linenomath}
could be placed. This improved bound~\eqref{eq:limit2008} based on
primary photons by a factor of $3000$
and proved the sensitivity of the new approach of testing secondary photons in air showers initiated by primary hadrons.

Still, an important limitation of bound~\eqref{eq:limit2017} is related to the uncertain composition of the primary cosmic rays. 
Due to this, most conservatively a pure proton composition had to be assumed.
As noted in~\cite{klinkhamer17}, this limitation could be overcome by including the shower-to-shower fluctutations $\sigma(X_\text{max})$ as an additional observable.
In contrast to $\left<X_\text{max}\right>$, the fluctuations show only a minor dependence on $\kappa$
(see also Fig.~\ref{fig:sigma_hadrons}). This may allow the exclusion of those composition assumptions that, for a given $\kappa$,
might be able to reproduce either $\left<X_\text{max}\right>$ or $\sigma(X_\text{max})$ alone, but not both observables simultaneously.

\section{Analysis}
\label{sec:analysis}

To analyze the impact of LV on the development of air showers, a full MC approach as in~\cite{klinkhamer17} is used.
The MC code CONEX~\cite{bergmann07a,pierog06a} was modified to include photon decay as well as the modified decay of the neutral pion.
Hadronic interactions are simulated with
EPOS LHC~\cite{pierog15a} and QGSJET-II-04~\cite{ostapchenko11a} using CONEX v2r5p40 as well as with SIBYLL~2.3d~\cite{riehn20a} using CONEX v2r7p50.
For all other settings, the defaults provided by the CONEX code are used.
We checked that the values derived from simulations performed with EPOS LHC and QGSJET-II-04 do not differ significantly between CONEX v2r5p40 and CONEX v2r7p50.

The exact composition of cosmic ray particles, especially at high energies, is unknown. To account for any possible composition of primary hadrons, four elements were chosen as representatives of their respective mass ranges. 
Chosen were protons (mass number $A$ = 1), helium nuclei ($A$ = 4), oxygen nuclei ($A$ = 16) and iron nuclei ($A$ = 56). 
The simulations performed for these different elements were then combined to simulate data taken from a set of air showers induced by different primary hadrons. 
A stepsize of $\unit[2]{\%}$ difference of the relative contributions of the individual elements between the different combinations was chosen.

\begin{figure}[tp]
\centering
\includegraphics[width=0.75\textwidth,]{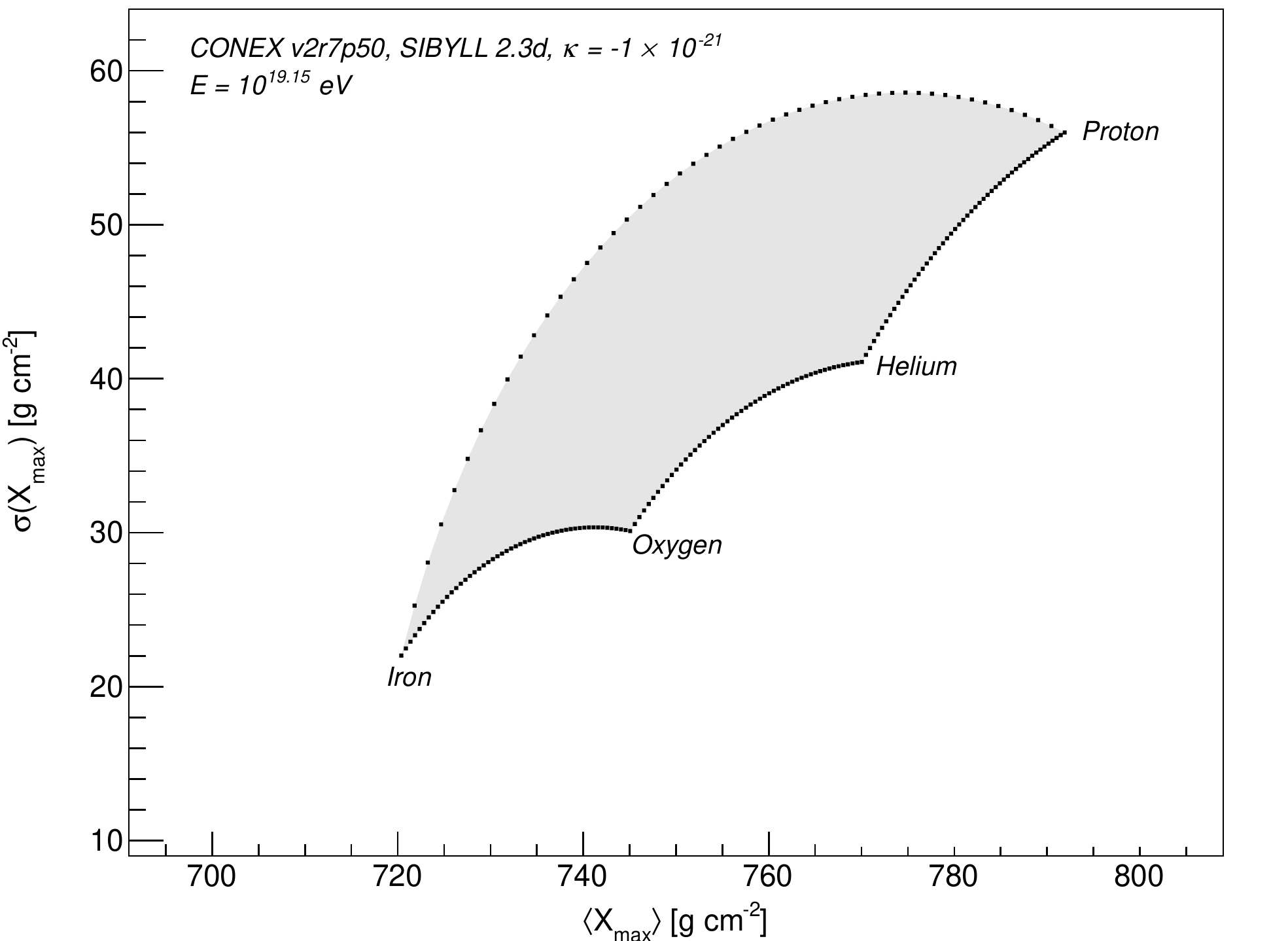}
\caption{The shaded region contains all possible values of $\left<X_\text{max}\right>$ and $\sigma(X_\text{max})$
for combinations of air showers induced by primary protons, helium, oxygen and iron nuclei
for $\kappa = -1 \times 10^{-21}$ and a primary particle energy of $\unit[10^{19.15}]{eV}$.
The ``edges'' refer to pure compositions as indicated.
Displayed are the proton-helium, helium-oxygen, oxygen-iron and iron-proton combinations.
Any point is differing $\unit[2]{\%}$ in composition from the neighboring points.
For instance, the upper curve resembles the iron-proton mixtures.
All other possible combinations produce
values of $\left<X_\text{max}\right>$ and $\sigma(X_\text{max})$ inside the umbrella-shaped area.} 
\label{fig:umbrella} 
\end{figure}

An example of the possible range of $\left<X_\text{max}\right>$ and $\sigma(X_\text{max})$ for fixed values of energy and $\kappa$ is displayed in Fig.~\ref{fig:umbrella}.
The well-known ``umbrella''-like shape (see, e.g.,~\cite{Kampert:2012mx}) is visible.
It should be kept in mind that
the resulting $\left<X_\text{max}\right>$ value of any combination is the same as the weighted mean of the $\left<X_\text{max}\right>$ values of all components.
In contrast, the $\sigma(X_\text{max})$ value of a set of showers with different primary hadrons is always greater than the weighted mean of the composites. 
This is due to the size of the shower-to-shower fluctuations increasing once showers induced by different particles with different mean shower depths are combined. 

The sets of simulated values obtained this way are then compared to the measurements taken by the Pierre Auger Observatory~\cite{auger19,auger14a}.
To accomplish a simultaneous comparison of both observables in the extended approach presented here, in each energy bin a two-dimensional confidence interval was used,
at a confidence level of $\unit[98]{\%}$ to have comparability to the previously derived limits. 
For this, the statistical and systematic uncertainties of the $\left<X_\text{max}\right>$ and $\sigma(X_\text{max})$ observations are approximated by Gaussian distributions (statistical) and uniform distributions (systematic)
and a contour line encompassing $\unit[98]{\%}$ of the distribution is drawn. The comparison is performed between all possible combinations of $\left<X_\text{max}\right>$ and $\sigma(X_\text{max})$ covered by the LV simulations and the Auger measurements. 

\begin{figure}[tp]
\centering
\includegraphics[width=0.75\textwidth,]{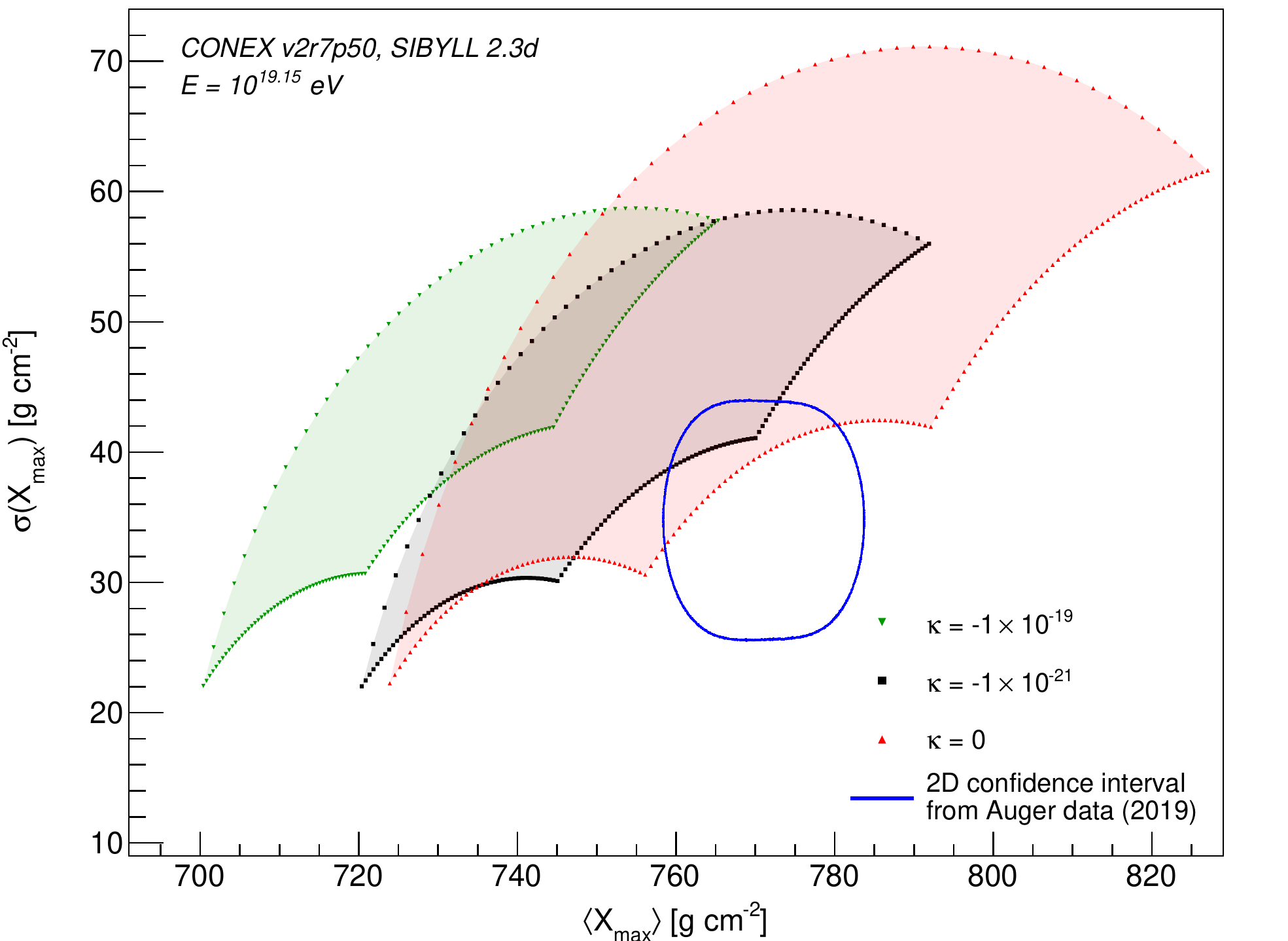}
\caption{Comparison of $\left<X_\text{max}\right>$ and $\sigma(X_\text{max})$ derived by simulations which incorporate LV to the 2D confidence intervals given by the measurements of the Pierre Auger Observatory~\cite{auger19,auger14a} for different values for $\kappa$ and a primary particle energy of $\unit[10^{19.15}]{eV}$.}
\label{fig:umbrella_kappa} 
\end{figure}

An illustration of such a comparison, as well as the change of $\left<X_\text{max}\right>$ and $\sigma(X_\text{max})$ in dependence on $\kappa$, can be seen in Fig.~\ref{fig:umbrella_kappa}. 
For  $\kappa = -1 \times 10^{-21}$, proton showers are significantly affected and iron showers only little, due to the smaller energy per nucleon.
Thus, compared to the case of $\kappa = 0$, the region of allowed values of $\left<X_\text{max}\right>$ and $\sigma(X_\text{max})$ shrinks considerably.
A further reduction of $\kappa$ to $\kappa = -1 \times 10^{-19}$ affects the different primaries in a more and more similar way.
Then, the main effect is a shift of the region towards smaller values of $\left<X_\text{max}\right>$ for decreasing values of $\kappa$.

\begin{figure}[tp]
\centering
\includegraphics[width=0.75\textwidth,]{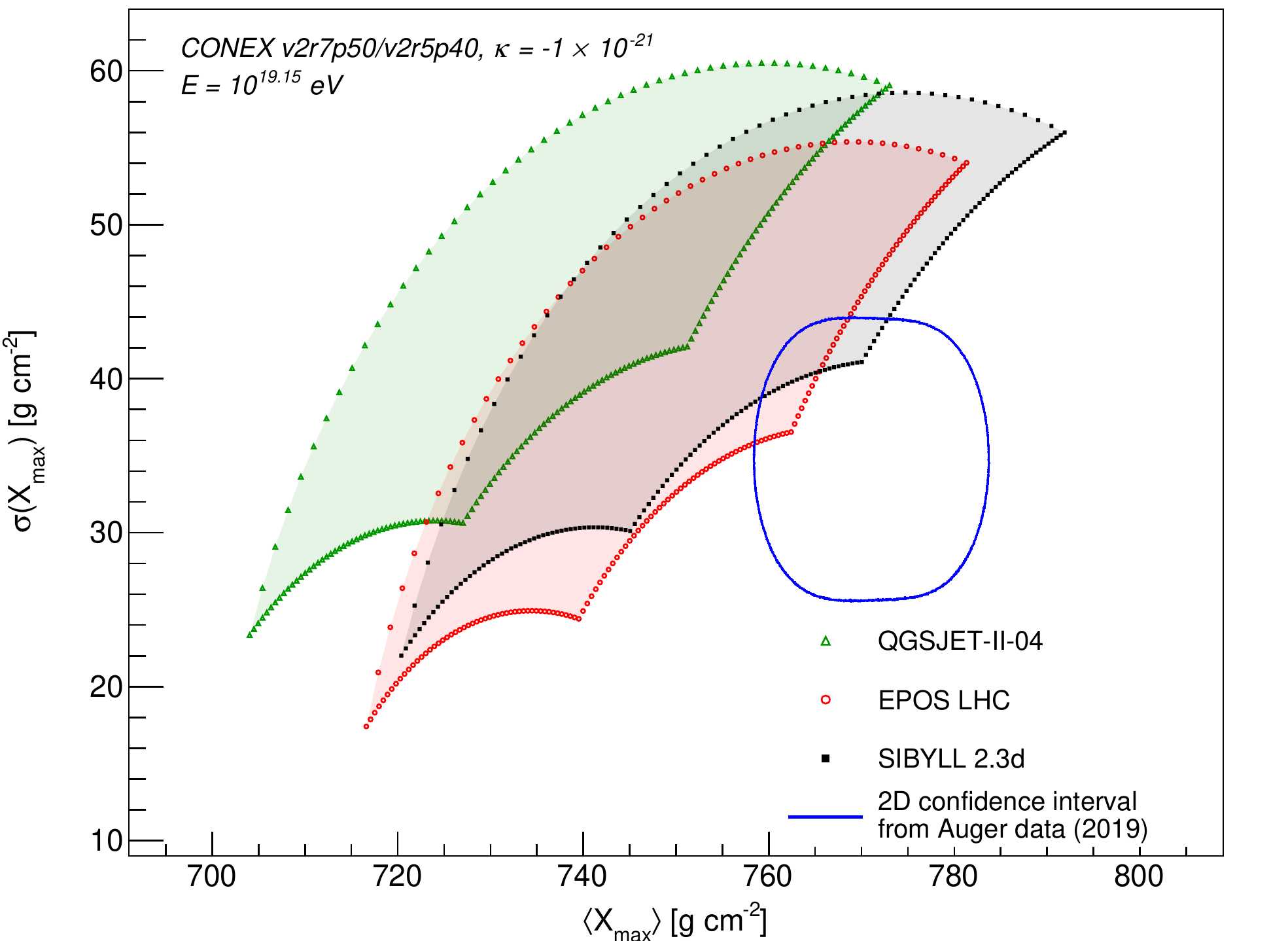}
\caption{Comparison of $\left<X_\text{max}\right>$ and $\sigma(X_\text{max})$ derived by simulations done with different hadronic interaction models which incorporate LV to the 2D confidence
  intervals given by the measurements of the Pierre Auger Observatory~\cite{auger19,auger14a} for $\kappa = -1 \times 10^{-21}$ and a primary particle energy of $\unit[10^{19.15}]{eV}$.}
\label{fig:umbrellaModels} 
\end{figure}

An overlap between two areas (simulated vs.\ observed) in the figure shows that there are primary hadron combinations which fit the Auger measurements.
Reversely, if for a specific value of $\kappa$ there is an energy at which no primary hadron combination fits the Auger measurements, 
it means this $\kappa$ does not fit the measurements and can thus be excluded. 
Scanning over $\kappa$ and the data energy bins, $\kappa_{\text{crit}}$ is found as the maximum value of $\kappa$ which can be excluded this way.
In other words, for $\kappa<\kappa_{\text{crit}}$, there is at least one energy bin where it is not possible to fit the measurements, whatever the primary hadron combination.
We excluded, for the time being, the two highest-energy bins (above $\unit[10^{19.50}]{eV}$) due to the comparably small statistics.

For the different hadronic interaction models this yields different values of $\kappa_{\text{crit}}$. An illustration of the differences between the values of $\left<X_\text{max}\right>$ and $\sigma(X_\text{max})$ for each model can be seen in  Fig.~\ref{fig:umbrellaModels}. The most conservative  $\kappa_{\text{crit}}$  is gained by using the SIBYLL~2.3d model which gives a new limit of
\begin{linenomath}
\begin{equation}
\kappa > \kappa_{\text{crit}} = -6 \times 10^{-21} ~~~  \text{($\unit[98]{\%}$ CL)} ~~~  \text{\small [SIBYLL~2.3d]}~.
\label{eq:newlimit}
\end{equation}
\end{linenomath}

A slightly stricter limit is achieved using the EPOS LHC model, resulting in a limit of $-5 \times 10^{-21}$. Due to the much shallower showers simulated with QGSJET-II-04, even for $\kappa = 0$ (no LV) the simulations are not able to reproduce the data in a self-consistent way.
This known fact (see e.g.~\cite{auger19}) indicates shortcomings in this specific hadronic interaction model.

The new bound of $\kappa_{\text{crit}} = -6 \times 10^{-21}$ improves
the previous bound~\eqref{eq:limit2017} by a factor of $50$. 

\begin{figure}[tp]
\centering
\includegraphics[width=0.75\textwidth,]{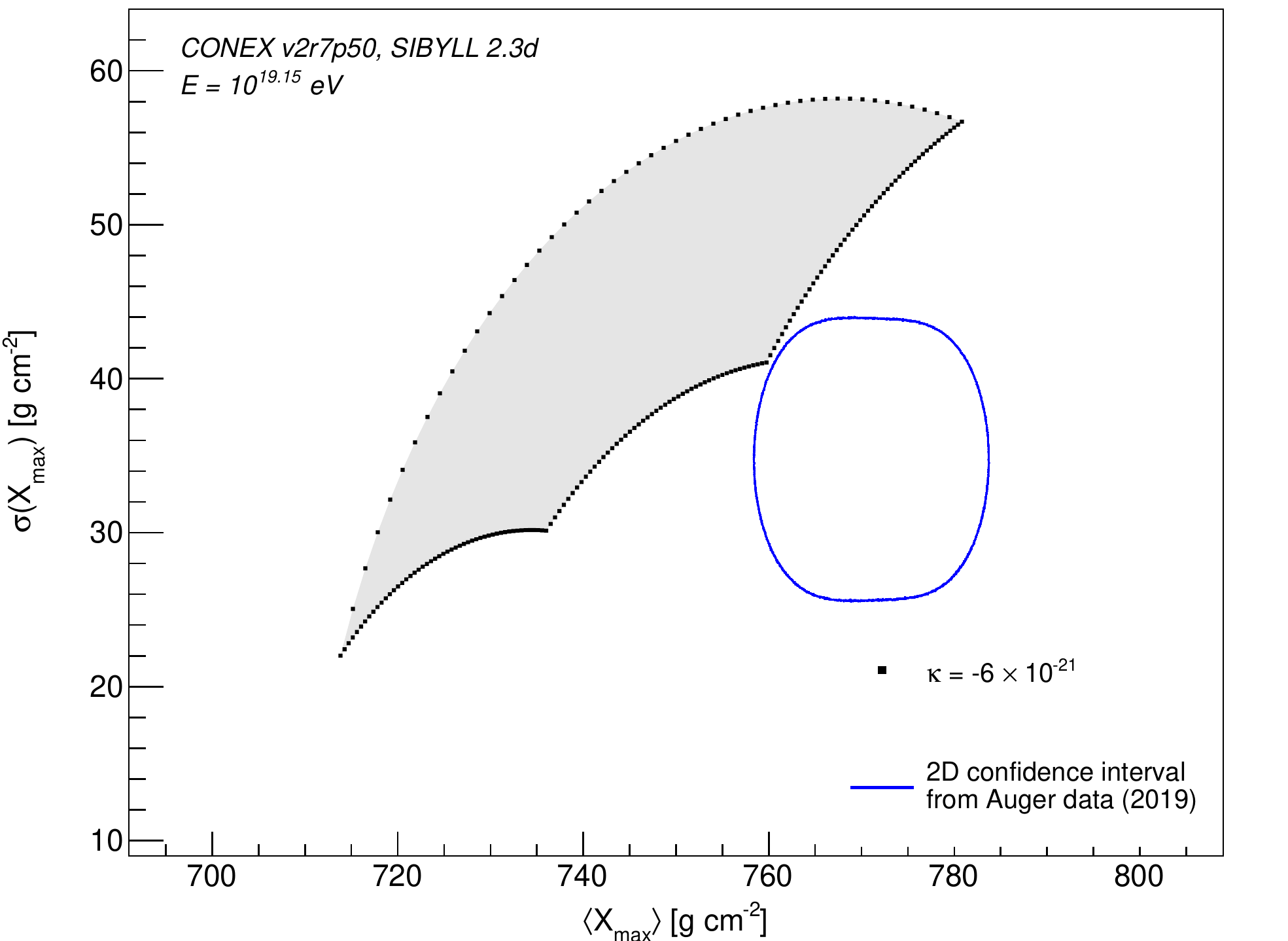}
\caption{Comparison of $\left<X_\text{max}\right>$ and $\sigma(X_\text{max})$ derived by LV simulations to the 2D confidence interval given by the measurements of the Pierre Auger Observatory for $\kappa_{\text{crit}} = -6 \times 10^{-21}$ and a primary particle energy of $\unit[10^{19.15}]{eV}$.}
\label{fig:umbrella_crit} 
\end{figure}

\section{Discussion}
\label{sec:discussion}
The energy bin driving the new limit in this paper is the energy range from $\unit[10^{19.1}]{eV}$ to $\unit[10^{19.2}]{eV}$ with a mean energy of $\unit[10^{19.15}]{eV}$.
This is primarily due to the observed $\sigma(X_\text{max})$ value being significantly lower than the one predicted for pure protons.
Only compositions with a fairly small contribution of protons are able to reproduce this observation.

In Fig.~\ref{fig:umbrella_crit}, all simulated combinations compared to the confidence interval derived from Auger data for this critical energy-$\kappa$ combination can be seen.
For this value of $\kappa$, the umbrella-shaped
area which encompasses all possible combinations of $\left<X_\text{max}\right>$ and $\sigma(X_\text{max})$ allowed by the LV simulations almost ``touches'' the range allowed by the Auger data.
With an increase in $\kappa$ (i.e., less strong LV) the value of $\left<X_\text{max}\right>$ also increases, which leads to a pure Helium composition being the CR-composition which first matches the experimental data. 
Further improvements of the bound can be expected when the possible compositions of primary cosmic ray particles can be further restricted. 

It is worth noting that an updated Auger data set is used in this work compared to the previous analysis~\cite{klinkhamer17} that led to bound~\eqref{eq:limit2017}. 
However, using the method detailed in~\cite{klinkhamer17}, based on $\left<X_\text{max}\right>$ alone,
would only yield minimal improvements in the previous bound on $\kappa$.
The main step forward here is the inclusion of $\sigma(X_\text{max})$ as a second observable.

The new bound is quite stable against the choice of the energy bin.
A limit of $-8 \times 10^{-21}$ would result from several other energy bins (in the range from $\unit[10^{18.8}]{eV}$ to $\unit[10^{19.1}]{eV}$).
Formally, the energy bin at $\unit[10^{19.55}]{eV}$ $-$ which we excluded here due to small statistics $-$ would yield a
somewhat stricter bound of  $-3 \times 10^{-21}$.

Further improvements on this bound can come from reduced experimental uncertainties.
For instance, uncertainties reduced by a factor of $2$ would lead in this case to bounds on $\kappa$ improved by more than an order of magnitude.
Stronger bounds also appear possible if additional observables, such as the signal size of the ground array, are taken into account.

In summary, we tested the presence of the decay of secondary UHE photons that are expected to be produced in extensive air showers.
Such decays, predicted as an LV effect in the theory framework considered,
can affect the longitudinal shower development in a significant and well-defined way.
Comparing to measurements by the Auger Observatory of $\left<X_\text{max}\right>$ and,
as a further observable added in this work,
the shower-to-shower fluctuation $\sigma(X_\text{max})$,
a new bound on the LV-parameter $\kappa$ was derived.
The new limit of $\kappa > -6 \times 10^{-21} ~  \text{($\unit[98]{\%}$ CL)} $
improves the previous bound by a factor of $50$. It should be noted
that in the theory considered here, LV is limited to the photon
sector. Through appropriate coordinate transformations, however, LV
can be moved to the fermion sector~\cite{Bailey:2004na,Altschul:2006zz}. Then,
bound~\ref{eq:newlimit} translates, in leading order, to
\begin{linenomath}
\begin{equation}
-\left[\kappa - (4/3)\,c_{00}^\text{e}\right] < 6 \times 10^{-21},
\label{eq:fermionlv}
\end{equation}
\end{linenomath}
where the coefficient $c_{00}^\text{e}$ denotes a possible
isotropic $c$-type LV in the fermion sector~\cite{Colladay:1998fq}.

Together with the present best limit on positive $\kappa$~\cite{klinkhamer08a,klinkhamer08b,klinkhamer08c} ,
where the mere existence of UHE cosmic rays was exploited
to exclude Vacuum Cherenkov radiation of the primary cosmic rays,
$\kappa$ is now bracketed by

\begin{linenomath}
\begin{equation}
6 \times 10^{-20} > \kappa > -6 \times 10^{-21} ~~~  \text{($\unit[98]{\%}$ CL)} ~.
\label{eq:kappabracket}
\end{equation}
\end{linenomath}

In this work, we focused on the effect of $\kappa < 0$ on the UHE shower development.
The corresponding analysis of the effect of $\kappa > 0$ will be reported on in a future study.

\subsection*{Acknowledgments}
The many fruitful discussions with Frans R. Klinkhamer are greatly appreciated.
We thank Tanguy Pierog for his help in modifying the CONEX source code.
This work was supported by the German Research Foundation (DFG project 408049454).

\bibliography{references}

\begin{thebibliography}{26}%
\makeatletter
\providecommand \@ifxundefined [1]{%
 \@ifx{#1\undefined}
}%
\providecommand \@ifnum [1]{%
 \ifnum #1\expandafter \@firstoftwo
 \else \expandafter \@secondoftwo
 \fi
}%
\providecommand \@ifx [1]{%
 \ifx #1\expandafter \@firstoftwo
 \else \expandafter \@secondoftwo
 \fi
}%
\providecommand \natexlab [1]{#1}%
\providecommand \enquote  [1]{``#1''}%
\providecommand \bibnamefont  [1]{#1}%
\providecommand \bibfnamefont [1]{#1}%
\providecommand \citenamefont [1]{#1}%
\providecommand \href@noop [0]{\@secondoftwo}%
\providecommand \href [0]{\begingroup \@sanitize@url \@href}%
\providecommand \@href[1]{\@@startlink{#1}\@@href}%
\providecommand \@@href[1]{\endgroup#1\@@endlink}%
\providecommand \@sanitize@url [0]{\catcode `\\12\catcode `\$12\catcode
  `\&12\catcode `\#12\catcode `\^12\catcode `\_12\catcode `\%12\relax}%
\providecommand \@@startlink[1]{}%
\providecommand \@@endlink[0]{}%
\providecommand \url  [0]{\begingroup\@sanitize@url \@url }%
\providecommand \@url [1]{\endgroup\@href {#1}{\urlprefix }}%
\providecommand \urlprefix  [0]{URL }%
\providecommand \Eprint [0]{\href }%
\providecommand \doibase [0]{http://dx.doi.org/}%
\providecommand \selectlanguage [0]{\@gobble}%
\providecommand \bibinfo  [0]{\@secondoftwo}%
\providecommand \bibfield  [0]{\@secondoftwo}%
\providecommand \translation [1]{[#1]}%
\providecommand \BibitemOpen [0]{}%
\providecommand \bibitemStop [0]{}%
\providecommand \bibitemNoStop [0]{.\EOS\space}%
\providecommand \EOS [0]{\spacefactor3000\relax}%
\providecommand \BibitemShut  [1]{\csname bibitem#1\endcsname}%
\let\auto@bib@innerbib\@empty
\bibitem [{\citenamefont {Liberati}\ and\ \citenamefont
  {Maccione}(2009)}]{liberati09a}%
  \BibitemOpen
  \bibfield  {author} {\bibinfo {author} {\bibfnamefont {S.}~\bibnamefont
  {Liberati}}\ and\ \bibinfo {author} {\bibfnamefont {L.}~\bibnamefont
  {Maccione}},\ }\href {\doibase 10.1146/annurev.nucl.010909.083640} {\bibfield
   {journal} {\bibinfo  {journal} {Ann. Rev. Nucl. Part. Sci.}\ }\textbf
  {\bibinfo {volume} {59}},\ \bibinfo {pages} {245} (\bibinfo {year} {2009})},\
  \Eprint {http://arxiv.org/abs/0906.0681} {arXiv:0906.0681 [astro-ph.HE]}
  \BibitemShut {NoStop}%
\bibitem [{\citenamefont {Kosteleck\'{y}}\ and\ \citenamefont
  {Russell}(2011)}]{kostelecky11a}%
  \BibitemOpen
  \bibfield  {author} {\bibinfo {author} {\bibfnamefont {V.~A.}\ \bibnamefont
  {Kosteleck\'{y}}}\ and\ \bibinfo {author} {\bibfnamefont {N.}~\bibnamefont
  {Russell}},\ }\href {\doibase 10.1103/RevModPhys.83.11} {\bibfield  {journal}
  {\bibinfo  {journal} {Rev. Mod. Phys.}\ }\textbf {\bibinfo {volume} {83}},\
  \bibinfo {pages} {11} (\bibinfo {year} {2011})},\ \Eprint
  {http://arxiv.org/abs/0801.0287} {regularly updated in arXiv:0801.0287
  [hep-ph]} \BibitemShut {NoStop}%
\bibitem [{\citenamefont {Klinkhamer}\ and\ \citenamefont
  {Risse}(2008{\natexlab{a}})}]{klinkhamer08a}%
  \BibitemOpen
  \bibfield  {author} {\bibinfo {author} {\bibfnamefont {F.~R.}\ \bibnamefont
  {Klinkhamer}}\ and\ \bibinfo {author} {\bibfnamefont {M.}~\bibnamefont
  {Risse}},\ }\href {\doibase 10.1103/PhysRevD.77.016002} {\bibfield  {journal}
  {\bibinfo  {journal} {Phys. Rev. D}\ }\textbf {\bibinfo {volume} {77}},\
  \bibinfo {pages} {016002} (\bibinfo {year} {2008}{\natexlab{a}})},\ \Eprint
  {http://arxiv.org/abs/0709.2502} {arXiv:0709.2502 [hep-ph]} \BibitemShut
  {NoStop}%
\bibitem [{\citenamefont {Klinkhamer}\ and\ \citenamefont
  {Risse}(2008{\natexlab{b}})}]{klinkhamer08b}%
  \BibitemOpen
  \bibfield  {author} {\bibinfo {author} {\bibfnamefont {F.~R.}\ \bibnamefont
  {Klinkhamer}}\ and\ \bibinfo {author} {\bibfnamefont {M.}~\bibnamefont
  {Risse}},\ }\href {\doibase 10.1103/PhysRevD.77.117901} {\bibfield  {journal}
  {\bibinfo  {journal} {Phys. Rev. D}\ }\textbf {\bibinfo {volume} {77}},\
  \bibinfo {pages} {117901} (\bibinfo {year} {2008}{\natexlab{b}})},\ \Eprint
  {http://arxiv.org/abs/0806.4351} {arXiv:0806.4351 [hep-ph]} \BibitemShut
  {NoStop}%
\bibitem [{\citenamefont {Klinkhamer}\ and\ \citenamefont
  {Schreck}(2008)}]{klinkhamer08c}%
  \BibitemOpen
  \bibfield  {author} {\bibinfo {author} {\bibfnamefont {F.~R.}\ \bibnamefont
  {Klinkhamer}}\ and\ \bibinfo {author} {\bibfnamefont {M.}~\bibnamefont
  {Schreck}},\ }\href {\doibase 10.1103/PhysRevD.78.085026} {\bibfield
  {journal} {\bibinfo  {journal} {Phys. Rev. D}\ }\textbf {\bibinfo {volume}
  {78}},\ \bibinfo {pages} {085026} (\bibinfo {year} {2008})},\ \Eprint
  {http://arxiv.org/abs/0809.3217} {arXiv:0809.3217 [hep-ph]} \BibitemShut
  {NoStop}%
\bibitem [{\citenamefont {Klinkhamer}\ \emph {et~al.}(2017)\citenamefont
  {Klinkhamer}, \citenamefont {Niechciol},\ and\ \citenamefont
  {Risse}}]{klinkhamer17}%
  \BibitemOpen
  \bibfield  {author} {\bibinfo {author} {\bibfnamefont {F.~R.}\ \bibnamefont
  {Klinkhamer}}, \bibinfo {author} {\bibfnamefont {M.}~\bibnamefont
  {Niechciol}}, \ and\ \bibinfo {author} {\bibfnamefont {M.}~\bibnamefont
  {Risse}},\ }\href {\doibase 10.1103/PhysRevD.96.116011} {\bibfield  {journal}
  {\bibinfo  {journal} {Phys. Rev. D}\ }\textbf {\bibinfo {volume} {96}},\
  \bibinfo {pages} {116011} (\bibinfo {year} {2017})},\ \Eprint
  {http://arxiv.org/abs/1710.02507} {arXiv:1710.02507 [hep-ph]} \BibitemShut
  {NoStop}%
\bibitem [{\citenamefont {Kosteleck\'{y}}\ and\ \citenamefont
  {Mewes}(2002)}]{kostelecky02a}%
  \BibitemOpen
  \bibfield  {author} {\bibinfo {author} {\bibfnamefont {V.~A.}\ \bibnamefont
  {Kosteleck\'{y}}}\ and\ \bibinfo {author} {\bibfnamefont {M.}~\bibnamefont
  {Mewes}},\ }\href {\doibase 10.1103/PhysRevD.66.056005} {\bibfield  {journal}
  {\bibinfo  {journal} {Phys. Rev. D}\ }\textbf {\bibinfo {volume} {66}},\
  \bibinfo {pages} {056005} (\bibinfo {year} {2002})},\ \Eprint
  {http://arxiv.org/abs/hep-ph/0205211} {arXiv:hep-ph/0205211} \BibitemShut
  {NoStop}%
\bibitem [{\citenamefont {D\'iaz}\ \emph {et~al.}(2016)\citenamefont {D\'iaz},
  \citenamefont {Klinkhamer},\ and\ \citenamefont {Risse}}]{diaz16a}%
  \BibitemOpen
  \bibfield  {author} {\bibinfo {author} {\bibfnamefont {J.~S.}\ \bibnamefont
  {D\'iaz}}, \bibinfo {author} {\bibfnamefont {F.~R.}\ \bibnamefont
  {Klinkhamer}}, \ and\ \bibinfo {author} {\bibfnamefont {M.}~\bibnamefont
  {Risse}},\ }\href {\doibase 10.1103/PhysRevD.94.085025} {\bibfield  {journal}
  {\bibinfo  {journal} {Phys. Rev. D}\ }\textbf {\bibinfo {volume} {94}},\
  \bibinfo {pages} {085025} (\bibinfo {year} {2016})},\ \Eprint
  {http://arxiv.org/abs/1607.02099} {arXiv:1607.02099 [hep-ph]} \BibitemShut
  {NoStop}%
\bibitem [{\citenamefont {Klinkhamer}(2018)}]{klinkhamer16a}%
  \BibitemOpen
  \bibfield  {author} {\bibinfo {author} {\bibfnamefont {F.~R.}\ \bibnamefont
  {Klinkhamer}},\ }\href {\doibase 10.1142/S0217732318501043} {\bibfield
  {journal} {\bibinfo  {journal} {Mod. Phys. Lett. A}\ }\textbf {\bibinfo
  {volume} {33}},\ \bibinfo {pages} {1850104} (\bibinfo {year} {2018})},\
  \Eprint {http://arxiv.org/abs/1610.03315} {arXiv:1610.03315 [hep-ph]}
  \BibitemShut {NoStop}%
\bibitem [{\citenamefont {Chadha}\ and\ \citenamefont
  {Nielsen}(1983)}]{chadha83a}%
  \BibitemOpen
  \bibfield  {author} {\bibinfo {author} {\bibfnamefont {S.}~\bibnamefont
  {Chadha}}\ and\ \bibinfo {author} {\bibfnamefont {H.~B.}\ \bibnamefont
  {Nielsen}},\ }\href {\doibase 10.1016/0550-3213(83)90081-0} {\bibfield
  {journal} {\bibinfo  {journal} {Nucl. Phys. B}\ }\textbf {\bibinfo {volume}
  {217}},\ \bibinfo {pages} {125} (\bibinfo {year} {1983})}\BibitemShut
  {NoStop}%
\bibitem [{\citenamefont {Colladay}\ and\ \citenamefont
  {Kostelecky}(1998)}]{Colladay:1998fq}%
  \BibitemOpen
  \bibfield  {author} {\bibinfo {author} {\bibfnamefont {D.}~\bibnamefont
  {Colladay}}\ and\ \bibinfo {author} {\bibfnamefont {V.~A.}\ \bibnamefont
  {Kostelecky}},\ }\href {\doibase 10.1103/PhysRevD.58.116002} {\bibfield
  {journal} {\bibinfo  {journal} {Phys. Rev. D}\ }\textbf {\bibinfo {volume}
  {58}},\ \bibinfo {pages} {116002} (\bibinfo {year} {1998})},\ \Eprint
  {http://arxiv.org/abs/hep-ph/9809521} {arXiv:hep-ph/9809521} \BibitemShut
  {NoStop}%
\bibitem [{\citenamefont {Klinkhamer}\ and\ \citenamefont
  {Schreck}(2011)}]{Klinkhamer:2010zs}%
  \BibitemOpen
  \bibfield  {author} {\bibinfo {author} {\bibfnamefont {F.~R.}\ \bibnamefont
  {Klinkhamer}}\ and\ \bibinfo {author} {\bibfnamefont {M.}~\bibnamefont
  {Schreck}},\ }\href {\doibase 10.1016/j.nuclphysb.2011.02.011} {\bibfield
  {journal} {\bibinfo  {journal} {Nucl. Phys. B}\ }\textbf {\bibinfo {volume}
  {848}},\ \bibinfo {pages} {90} (\bibinfo {year} {2011})},\ \Eprint
  {http://arxiv.org/abs/1011.4258} {arXiv:1011.4258 [hep-th]} \BibitemShut
  {NoStop}%
\bibitem [{\citenamefont {Bernadotte}\ and\ \citenamefont
  {Klinkhamer}(2007)}]{Bernadotte:2006ya}%
  \BibitemOpen
  \bibfield  {author} {\bibinfo {author} {\bibfnamefont {S.}~\bibnamefont
  {Bernadotte}}\ and\ \bibinfo {author} {\bibfnamefont {F.~R.}\ \bibnamefont
  {Klinkhamer}},\ }\href {\doibase 10.1103/PhysRevD.75.024028} {\bibfield
  {journal} {\bibinfo  {journal} {Phys. Rev. D}\ }\textbf {\bibinfo {volume}
  {75}},\ \bibinfo {pages} {024028} (\bibinfo {year} {2007})},\ \Eprint
  {http://arxiv.org/abs/hep-ph/0610216} {arXiv:hep-ph/0610216} \BibitemShut
  {NoStop}%
\bibitem [{\citenamefont {Klinkhamer}\ and\ \citenamefont
  {Schreck}(2012)}]{Klinkhamer:2011ez}%
  \BibitemOpen
  \bibfield  {author} {\bibinfo {author} {\bibfnamefont {F.~R.}\ \bibnamefont
  {Klinkhamer}}\ and\ \bibinfo {author} {\bibfnamefont {M.}~\bibnamefont
  {Schreck}},\ }\href {\doibase 10.1016/j.nuclphysb.2011.11.019} {\bibfield
  {journal} {\bibinfo  {journal} {Nucl. Phys. B}\ }\textbf {\bibinfo {volume}
  {856}},\ \bibinfo {pages} {666} (\bibinfo {year} {2012})},\ \Eprint
  {http://arxiv.org/abs/1110.4101} {arXiv:1110.4101 [hep-th]} \BibitemShut
  {NoStop}%
\bibitem [{\citenamefont {D\'iaz}\ and\ \citenamefont
  {Klinkhamer}(2015)}]{diaz15a}%
  \BibitemOpen
  \bibfield  {author} {\bibinfo {author} {\bibfnamefont {J.~S.}\ \bibnamefont
  {D\'iaz}}\ and\ \bibinfo {author} {\bibfnamefont {F.~R.}\ \bibnamefont
  {Klinkhamer}},\ }\href {\doibase 10.1103/PhysRevD.92.025007} {\bibfield
  {journal} {\bibinfo  {journal} {Phys. Rev. D}\ }\textbf {\bibinfo {volume}
  {92}},\ \bibinfo {pages} {025007} (\bibinfo {year} {2015})},\ \Eprint
  {http://arxiv.org/abs/1504.01324} {arXiv:1504.01324 [hep-ph]} \BibitemShut
  {NoStop}%
\bibitem [{\citenamefont {{Niechciol for the Pierre Auger
  Collaboration}}(2017)}]{niechciol17a}%
  \BibitemOpen
  \bibfield  {author} {\bibinfo {author} {\bibfnamefont {M.}~\bibnamefont
  {{Niechciol for the Pierre Auger Collaboration}}},\ }\href@noop {} {\bibfield
   {journal} {\bibinfo  {journal} {PoS}\ }\textbf {\bibinfo {volume}
  {ICRC2017}},\ \bibinfo {pages} {517} (\bibinfo {year} {2017})},\ \Eprint
  {http://arxiv.org/abs/1708.06592} {arXiv:1708.06592 [astro-ph.HE]}
  \BibitemShut {NoStop}%
\bibitem [{\citenamefont {{Yushkov for the Pierre Auger
  Collaboration}}(2019)}]{auger19}%
  \BibitemOpen
  \bibfield  {author} {\bibinfo {author} {\bibfnamefont {A.}~\bibnamefont
  {{Yushkov for the Pierre Auger Collaboration}}},\ }\href@noop {} {\bibfield
  {journal} {\bibinfo  {journal} {PoS}\ }\textbf {\bibinfo {volume}
  {ICRC2019}},\ \bibinfo {pages} {482} (\bibinfo {year} {2019})},\ \Eprint
  {http://arxiv.org/abs/1909.09073} {arXiv:1909.09073 [astro-ph.HE]}
  \BibitemShut {NoStop}%
\bibitem [{\citenamefont {{The Pierre Auger Collaboration}}(2014)}]{auger14a}%
  \BibitemOpen
  \bibfield  {author} {\bibinfo {author} {\bibnamefont {{The Pierre Auger
  Collaboration}}},\ }\href {\doibase 10.1103/PhysRevD.90.122005} {\bibfield
  {journal} {\bibinfo  {journal} {Phys. Rev. D}\ }\textbf {\bibinfo {volume}
  {90}},\ \bibinfo {pages} {122005} (\bibinfo {year} {2014})},\ \Eprint
  {http://arxiv.org/abs/1409.4809} {arXiv:1409.4809 [astro-ph.HE]} \BibitemShut
  {NoStop}%
\bibitem [{\citenamefont {Bergmann}\ \emph {et~al.}(2007)\citenamefont
  {Bergmann}, \citenamefont {Engel}, \citenamefont {Heck}, \citenamefont
  {Kalmykov}, \citenamefont {Ostapchenko}, \citenamefont {Pierog},
  \citenamefont {Thouw},\ and\ \citenamefont {Werner}}]{bergmann07a}%
  \BibitemOpen
  \bibfield  {author} {\bibinfo {author} {\bibfnamefont {T.}~\bibnamefont
  {Bergmann}}, \bibinfo {author} {\bibfnamefont {R.}~\bibnamefont {Engel}},
  \bibinfo {author} {\bibfnamefont {D.}~\bibnamefont {Heck}}, \bibinfo {author}
  {\bibfnamefont {N.~N.}\ \bibnamefont {Kalmykov}}, \bibinfo {author}
  {\bibfnamefont {S.}~\bibnamefont {Ostapchenko}}, \bibinfo {author}
  {\bibfnamefont {T.}~\bibnamefont {Pierog}}, \bibinfo {author} {\bibfnamefont
  {T.}~\bibnamefont {Thouw}}, \ and\ \bibinfo {author} {\bibfnamefont
  {K.}~\bibnamefont {Werner}},\ }\href {\doibase
  10.1016/j.astropartphys.2006.08.005} {\bibfield  {journal} {\bibinfo
  {journal} {Astropart. Phys.}\ }\textbf {\bibinfo {volume} {26}},\ \bibinfo
  {pages} {420} (\bibinfo {year} {2007})},\ \Eprint
  {http://arxiv.org/abs/astro-ph/0606564} {arXiv:astro-ph/0606564} \BibitemShut
  {NoStop}%
\bibitem [{\citenamefont {Pierog}\ \emph {et~al.}(2006)\citenamefont {Pierog},
  \citenamefont {Alekseeva}, \citenamefont {Bergmann}, \citenamefont
  {Chernatkin}, \citenamefont {Engel}, \citenamefont {Heck}, \citenamefont
  {Kalmykov}, \citenamefont {Moyon}, \citenamefont {Ostapchenko}, \citenamefont
  {Thouw},\ and\ \citenamefont {Werner}}]{pierog06a}%
  \BibitemOpen
  \bibfield  {author} {\bibinfo {author} {\bibfnamefont {T.}~\bibnamefont
  {Pierog}}, \bibinfo {author} {\bibfnamefont {M.~K.}\ \bibnamefont
  {Alekseeva}}, \bibinfo {author} {\bibfnamefont {T.}~\bibnamefont {Bergmann}},
  \bibinfo {author} {\bibfnamefont {V.}~\bibnamefont {Chernatkin}}, \bibinfo
  {author} {\bibfnamefont {R.}~\bibnamefont {Engel}}, \bibinfo {author}
  {\bibfnamefont {D.}~\bibnamefont {Heck}}, \bibinfo {author} {\bibfnamefont
  {N.~N.}\ \bibnamefont {Kalmykov}}, \bibinfo {author} {\bibfnamefont
  {J.}~\bibnamefont {Moyon}}, \bibinfo {author} {\bibfnamefont
  {S.}~\bibnamefont {Ostapchenko}}, \bibinfo {author} {\bibfnamefont
  {T.}~\bibnamefont {Thouw}}, \ and\ \bibinfo {author} {\bibfnamefont
  {K.}~\bibnamefont {Werner}},\ }\href {\doibase
  10.1016/j.nuclphysbps.2005.07.029} {\bibfield  {journal} {\bibinfo  {journal}
  {Nucl. Phys. Proc. Suppl.}\ }\textbf {\bibinfo {volume} {151}},\ \bibinfo
  {pages} {159} (\bibinfo {year} {2006})},\ \Eprint
  {http://arxiv.org/abs/astro-ph/0411260} {arXiv:astro-ph/0411260} \BibitemShut
  {NoStop}%
\bibitem [{\citenamefont {Pierog}\ \emph {et~al.}(2015)\citenamefont {Pierog},
  \citenamefont {Karpenko}, \citenamefont {Katzy}, \citenamefont {Yatsenko},\
  and\ \citenamefont {Werner}}]{pierog15a}%
  \BibitemOpen
  \bibfield  {author} {\bibinfo {author} {\bibfnamefont {T.}~\bibnamefont
  {Pierog}}, \bibinfo {author} {\bibfnamefont {I.}~\bibnamefont {Karpenko}},
  \bibinfo {author} {\bibfnamefont {J.~M.}\ \bibnamefont {Katzy}}, \bibinfo
  {author} {\bibfnamefont {E.}~\bibnamefont {Yatsenko}}, \ and\ \bibinfo
  {author} {\bibfnamefont {K.}~\bibnamefont {Werner}},\ }\href {\doibase
  10.1103/PhysRevC.92.034906} {\bibfield  {journal} {\bibinfo  {journal} {Phys.
  Rev. C}\ }\textbf {\bibinfo {volume} {92}},\ \bibinfo {pages} {034906}
  (\bibinfo {year} {2015})},\ \Eprint {http://arxiv.org/abs/1306.0121}
  {arXiv:1306.0121 [hep-ph]} \BibitemShut {NoStop}%
\bibitem [{\citenamefont {Ostapchenko}(2011)}]{ostapchenko11a}%
  \BibitemOpen
  \bibfield  {author} {\bibinfo {author} {\bibfnamefont {S.}~\bibnamefont
  {Ostapchenko}},\ }\href {\doibase 10.1103/PhysRevD.83.014018} {\bibfield
  {journal} {\bibinfo  {journal} {Physical Review D}\ }\textbf {\bibinfo
  {volume} {83}},\ \bibinfo {pages} {014018} (\bibinfo {year} {2011})},\
  \Eprint {http://arxiv.org/abs/1010.1869} {arXiv:1010.1869 [hep-ph]}
  \BibitemShut {NoStop}%
\bibitem [{\citenamefont {Riehn}\ \emph {et~al.}(2020)\citenamefont {Riehn},
  \citenamefont {Engel}, \citenamefont {Fedynitch}, \citenamefont {Gaisser},\
  and\ \citenamefont {Stanev}}]{riehn20a}%
  \BibitemOpen
  \bibfield  {author} {\bibinfo {author} {\bibfnamefont {F.}~\bibnamefont
  {Riehn}}, \bibinfo {author} {\bibfnamefont {R.}~\bibnamefont {Engel}},
  \bibinfo {author} {\bibfnamefont {A.}~\bibnamefont {Fedynitch}}, \bibinfo
  {author} {\bibfnamefont {T.~K.}\ \bibnamefont {Gaisser}}, \ and\ \bibinfo
  {author} {\bibfnamefont {T.}~\bibnamefont {Stanev}},\ }\href {\doibase
  10.1103/PhysRevD.102.063002} {\bibfield  {journal} {\bibinfo  {journal}
  {Phys. Rev. D}\ }\textbf {\bibinfo {volume} {102}},\ \bibinfo {pages}
  {063002} (\bibinfo {year} {2020})},\ \Eprint
  {http://arxiv.org/abs/1912.03300} {arXiv:1912.03300 [hep-ph]} \BibitemShut
  {NoStop}%
\bibitem [{\citenamefont {Kampert}\ and\ \citenamefont
  {Unger}(2012)}]{Kampert:2012mx}%
  \BibitemOpen
  \bibfield  {author} {\bibinfo {author} {\bibfnamefont {K.-H.}\ \bibnamefont
  {Kampert}}\ and\ \bibinfo {author} {\bibfnamefont {M.}~\bibnamefont
  {Unger}},\ }\href {\doibase 10.1016/j.astropartphys.2012.02.004} {\bibfield
  {journal} {\bibinfo  {journal} {Astropart. Phys.}\ }\textbf {\bibinfo
  {volume} {35}},\ \bibinfo {pages} {660} (\bibinfo {year} {2012})},\ \Eprint
  {http://arxiv.org/abs/1201.0018} {arXiv:1201.0018 [astro-ph.HE]} \BibitemShut
  {NoStop}%
\bibitem [{\citenamefont {Bailey}\ and\ \citenamefont
  {Kostelecky}(2004)}]{Bailey:2004na}%
  \BibitemOpen
  \bibfield  {author} {\bibinfo {author} {\bibfnamefont {Q.~G.}\ \bibnamefont
  {Bailey}}\ and\ \bibinfo {author} {\bibfnamefont {V.~A.}\ \bibnamefont
  {Kostelecky}},\ }\href {\doibase 10.1103/PhysRevD.70.076006} {\bibfield
  {journal} {\bibinfo  {journal} {Phys. Rev. D}\ }\textbf {\bibinfo {volume}
  {70}},\ \bibinfo {pages} {076006} (\bibinfo {year} {2004})},\ \Eprint
  {http://arxiv.org/abs/hep-ph/0407252} {arXiv:hep-ph/0407252} \BibitemShut
  {NoStop}%
\bibitem [{\citenamefont {Altschul}(2007)}]{Altschul:2006zz}%
  \BibitemOpen
  \bibfield  {author} {\bibinfo {author} {\bibfnamefont {B.}~\bibnamefont
  {Altschul}},\ }\href {\doibase 10.1103/PhysRevLett.98.041603} {\bibfield
  {journal} {\bibinfo  {journal} {Phys. Rev. Lett.}\ }\textbf {\bibinfo
  {volume} {98}},\ \bibinfo {pages} {041603} (\bibinfo {year} {2007})},\
  \Eprint {http://arxiv.org/abs/hep-th/0609030} {arXiv:hep-th/0609030}
  \BibitemShut {NoStop}%
\end{thebibliography}%

\end{document}